\documentclass[a4paper]{article}

\usepackage{INTERSPEECH2021}
\usepackage[colorlinks,hidelinks=1]{hyperref}
\usepackage{multicol}  
\usepackage{multirow}  
\usepackage{graphicx}
\usepackage{subfigure}
\title{DBNet: A Dual-branch Network Architecture Processing on Spectrum and Waveform for Single-channel Speech Enhancement }
\name{Kanghao Zhang, Shulin He, Hao Li, Xueliang Zhang}
\address{College of Computer Science, Inner Mongolia University, China}
\email{\{imu.koer, heshulin, lihao\}@mail.imu.edu.cn, cszxl@imu.edu.cn}

\begin{document}

\maketitle
\begin{abstract}
    In real acoustic environment, speech enhancement is an arduous task to improve the quality and intelligibility of speech interfered by background noise and reverberation. Over the past years, deep learning has shown great potential on speech enhancement. In this paper, we propose a novel real-time framework called DBNet which is a dual-branch structure with alternate interconnection. Each branch incorporates an encoder-decoder architecture with skip connections. The two branches are responsible for spectrum and waveform  modeling, respectively. A bridge layer is adopted to exchange information between the two branches. Systematic evaluation and comparison show that the proposed system substantially outperforms related algorithms under very challenging environments. And in INTERSPEECH 2021 Deep Noise Suppression (DNS) challenge, the proposed system ranks the top 8 in real-time track 1 in terms of the Mean Opinion Score (MOS) of the ITU-T P.835 framework.
\end{abstract}
\noindent\textbf{Index Terms}: Speech enhancement, time domain, frequency domain, real-time

\section{Introduction}
    With the COVID-19 pandemic, lots people are working online, and the demand for reliable real-time speech enhancement algorithms has increased sharply. During these times, we need to ensure clear call quality and effective communication and cooperation with others without delay. Our communication are usually disturbed by a lot of background noise, such as washing machines, passing trucks, and the strong reverberation. These will affect the efficiency of our work and communication. Recently, many researchers from academia and industry have made significant contributions to monaural speech enhancement. However, due to the diversity of noise types in reality, even the the state-of-the-art algorithms cannot handle challenging environments well.

    With the development of deep learning, many research regard speech enhancement as a supervised learning problem ~\cite{pandey2019new}~\cite{tan2019learning}~\cite{wang2018supervised}, and obtained excellent performance. Usually the neural network input is time domain signal or STFT domain signal. In ~\cite{tan2019learning}~\cite{li2021real}, the authors study the noise reduction problem in the short-time Fourier transform (STFT) domain. In ~\cite{pandey2019new}~\cite{pandey2020densely}, the time-domain signal after framing is directly feed to the neural network. Both frequency domain and time domain has its own advantages. The STFT method is more in line with human hearing perception, and the characteristics of speech is more explicit. Time domain method does not damage the signal by STFT, avoids the well-known invalid STFT problem.

    Lim et al. ~\cite{lim2018time} introduced a time-frequency network to jointly optimize the time and frequency domains of a signal for the task of audio super resolution. They show that combine these two domains can boost the audio super resolution performance and obtain the state-of-the-art in both quantitatively and qualitatively.

\begin{figure}[h] 
\centering
\subfigure[Time domain noisy with an impulse-like noise]{\label{subfig:wav_impulse} \includegraphics[width=3.5cm]{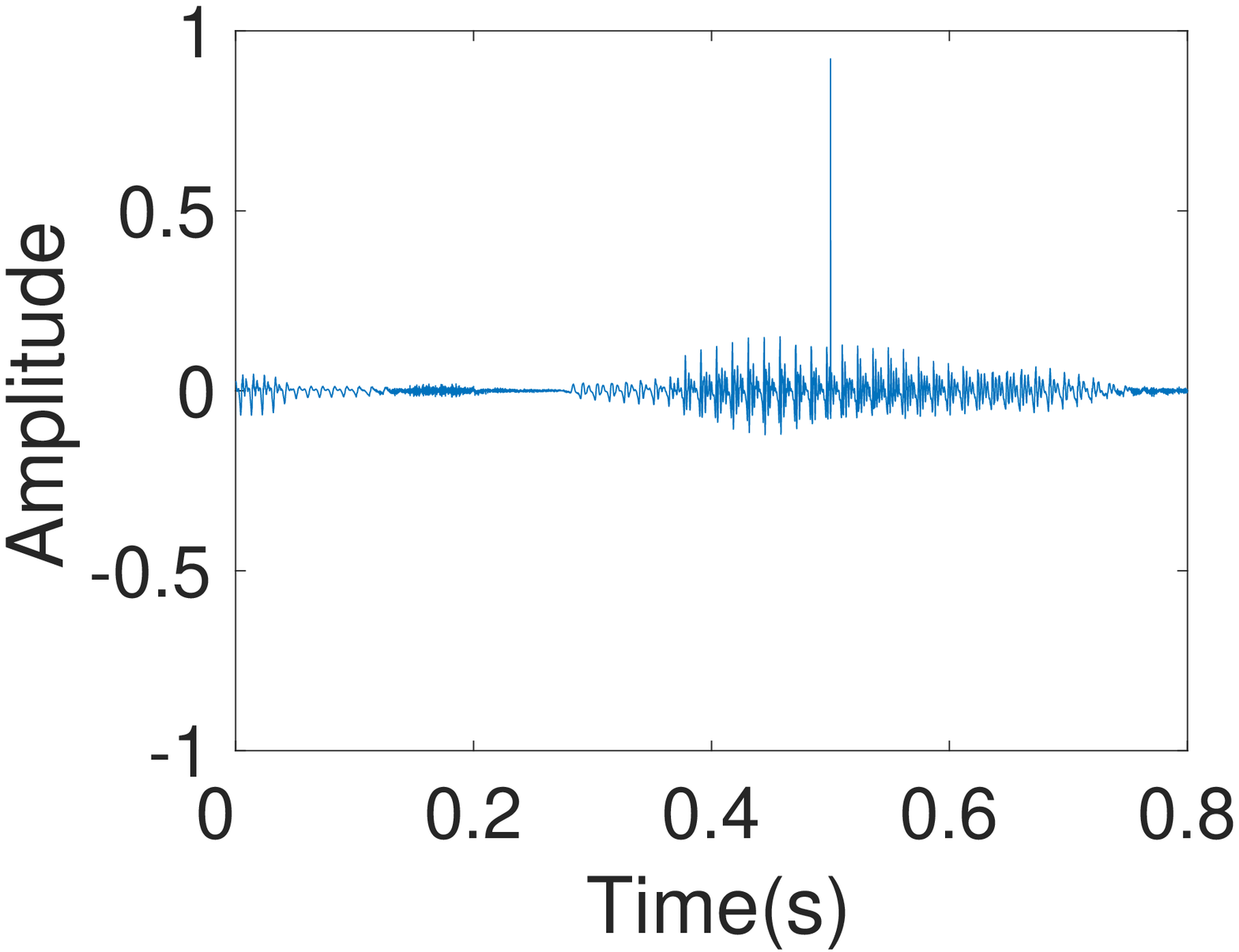}} 
\subfigure[Frequency domain noisy with an impulse-like noise]{\label{subfig:mag_impulse} \includegraphics[width=3.5cm]{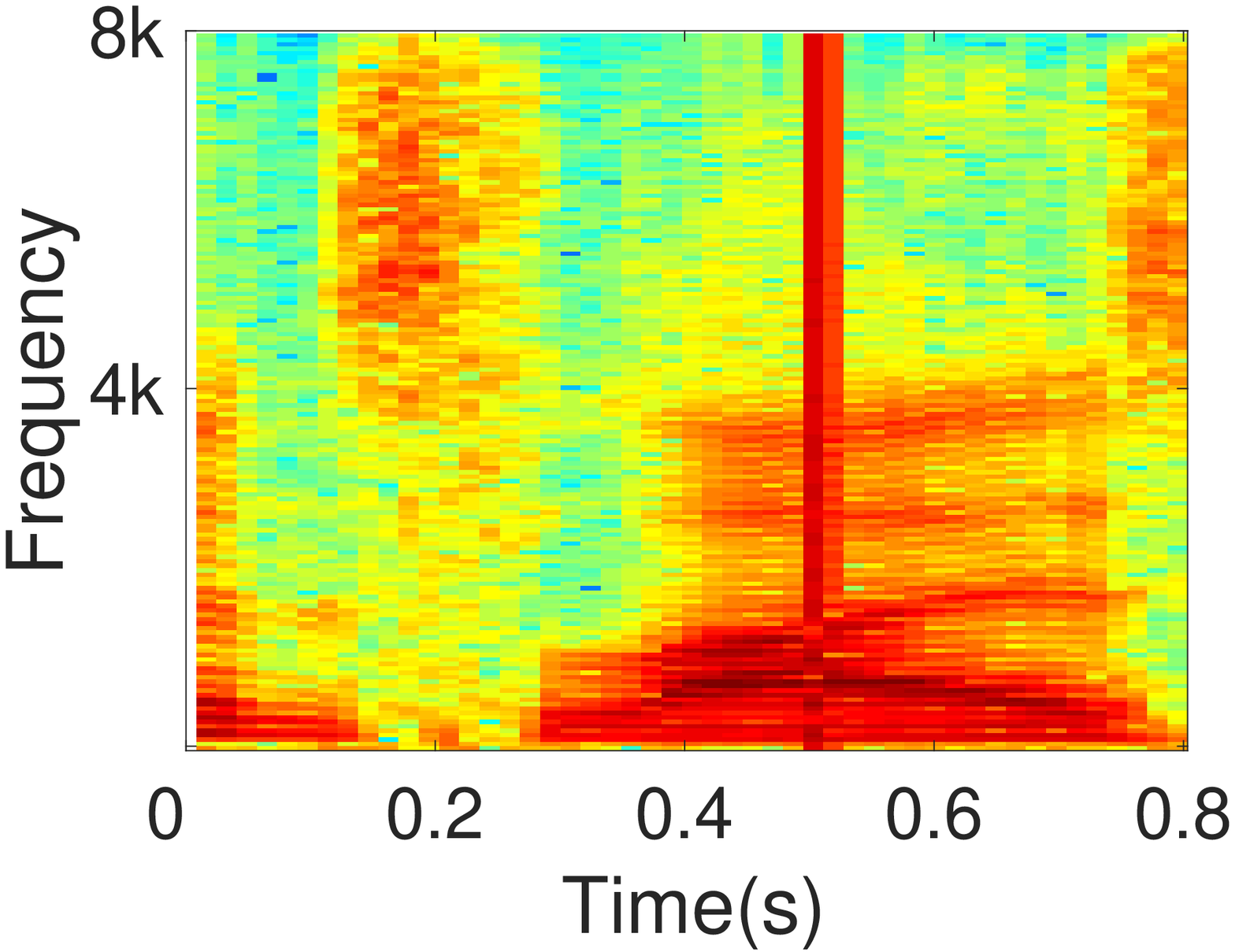}} 
\subfigure[Time domain noisy with a narrowband-like noise]{\label{subfig:wav_sin} \includegraphics[width=3.5cm]{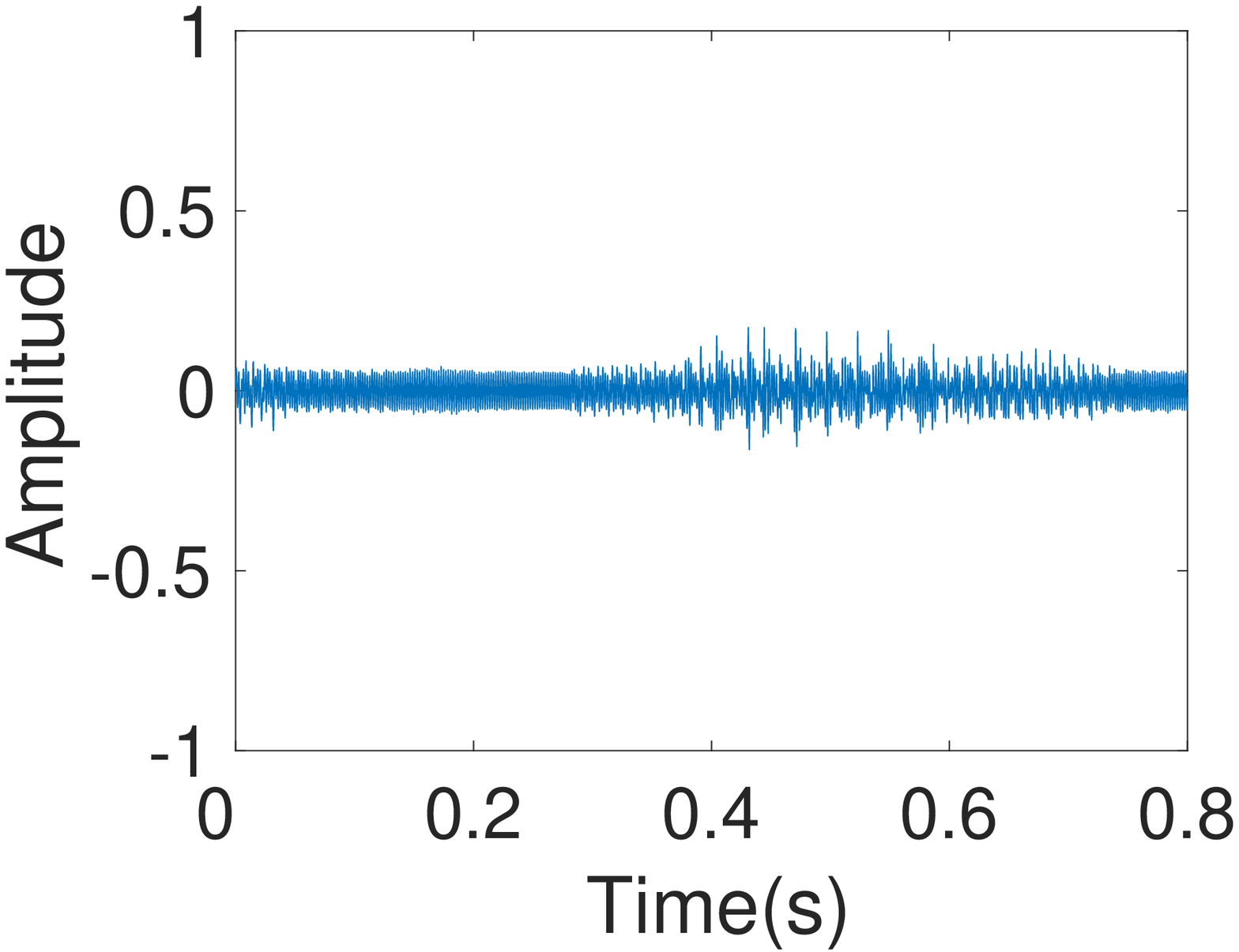}} 
\subfigure[Frequency domain noisy with a narrowband-like noise]{\label{subfig:mag_sin}  \includegraphics[width=3.5cm]{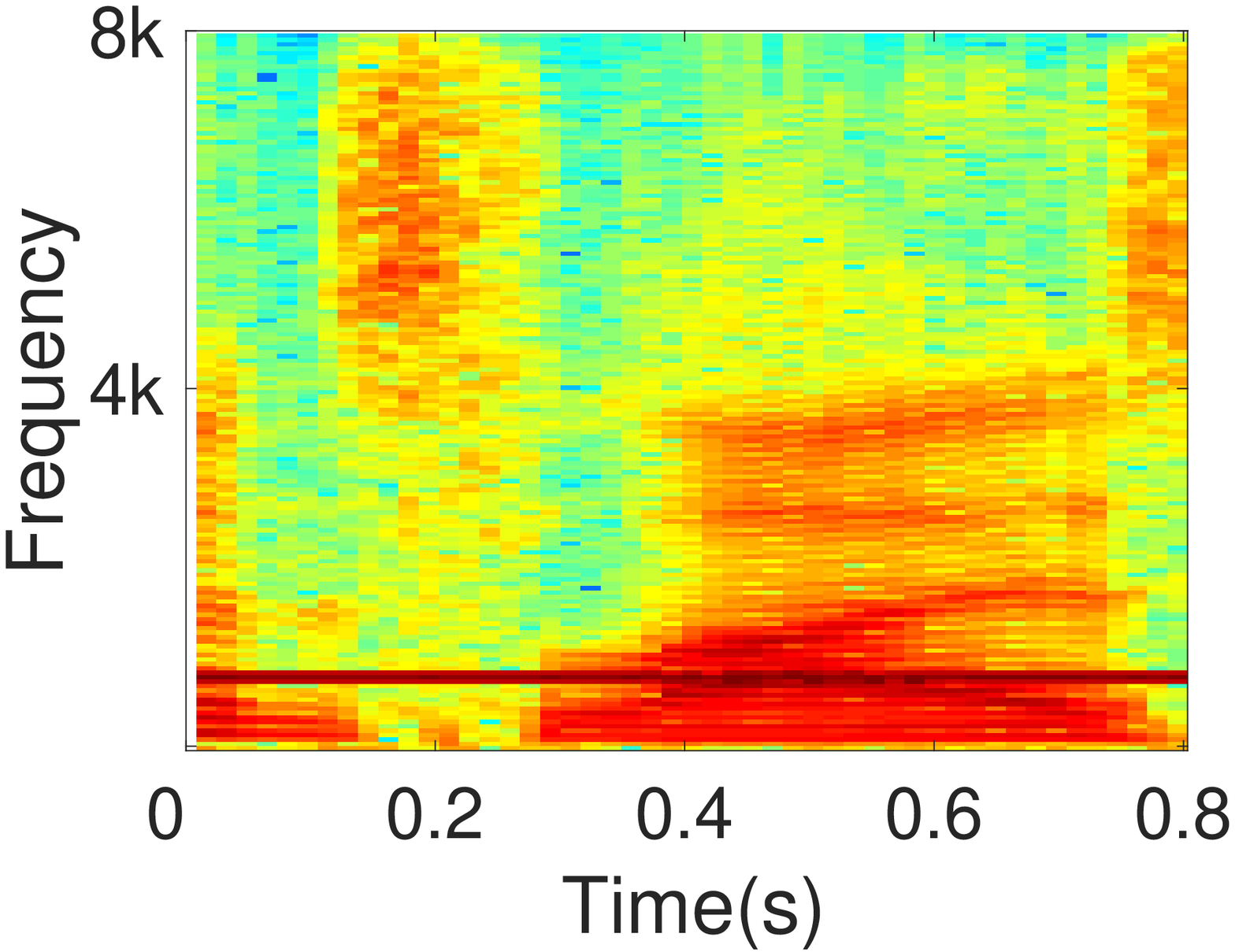}} 

\caption{Time- and frequency-domain noisy signal with two kinds of noises.}
\label{fig:noise_type}
\end{figure}

    As we know, for an impulse-like noise as shown in Fig. \ref{subfig:wav_impulse} and \ref{subfig:mag_impulse}, it is easy to eliminate in the time domain. Only a few samples need to be removed. In the frequency domain, the impulse-like nose pollutes the entire frequency band, and it is difficult to be eliminated by a speech enhancement method based on the frequency domain.
    In contrast, for a narrowband-like noise as shown in Figure ~\ref{subfig:wav_sin} and ~\ref{subfig:mag_sin}, the noise is distributed on the narrowband and the frequency domain-based method covers well. In the time-domain, the noise and the speech are coupled together, and it is hard to decouple by an enhancement method based on time-domain.
    In this paper, in order to reduce noise better, we proposed a novel speech enhancement algorithm called DBNet, which combines the time domain and frequency domain together.
    The DBNet is a dual-branch structure with alternate interconnection. Each branch incorporates an encoder decoder architecture with skip connections. The two branches are responsible for spectrum and waveform modeling, respectively. A bridge layer is adopted to exchange information between the two branches. Experiments show that the proposed method has excellent results on the WSJ0 SI-84 ~\cite{WSJ0} and DNS Challenge ~\cite{reddy2021interspeech}.

    The rest of the paper is organized as follows. The structure of the proposed DBNet is described in Section \ref{Sec:model descripthon}. Section \ref{Sec:experiment} describes the experimental setup. The experimental results are revealed in Section \ref{Sec:result}. We conclude this paper in Section \ref{Sec:conclusions}.

\begin{figure*}[!ht]
  \centering
  \includegraphics[width=0.7\textwidth]{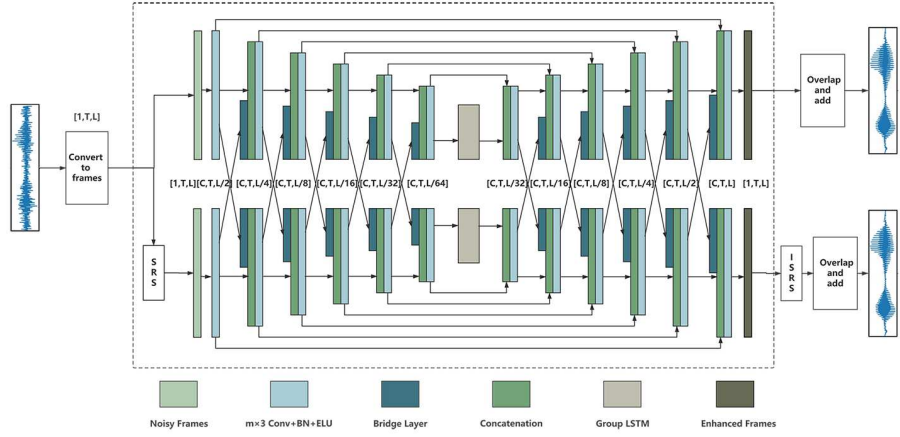}
  \caption{Diagram of the proposed DBNet model}
  \label{fig:overview}
\end{figure*} 
\section{Model Description}
\label{Sec:model descripthon}
    The overall structure of the proposed method is shown in Fig. \ref{fig:overview}. In the following sections, three important parts: SRS module, GLSTM, bridge layer, will be introduced one by one.

\subsection{Shift Real Spectral (SRS) Module }
    Recently, Liu et al.~\cite{SRS} proposed a new separation target based on the time-frequency representation of SRS in their work, which proved the superiority of SRS over STFT. First, SRS takes the phase into consideration, which improves the speech intelligibility and quality. Second, SRS is a spectral representation method in real field instead of the complex field, all of the elements of input are real numbers. So it reduces the modeling difficulty and provide convenience for the information interaction module of our model. Therefore in this paper, SRS is adopted as our frequency domain input based on the two advantages above.

\subsection{Gated Convolution and Group LSTM}
\begin{figure}[h] 
\centering
\subfigure[Gated Convolution]{
\label{fig:a}
\includegraphics[width=1.3in]{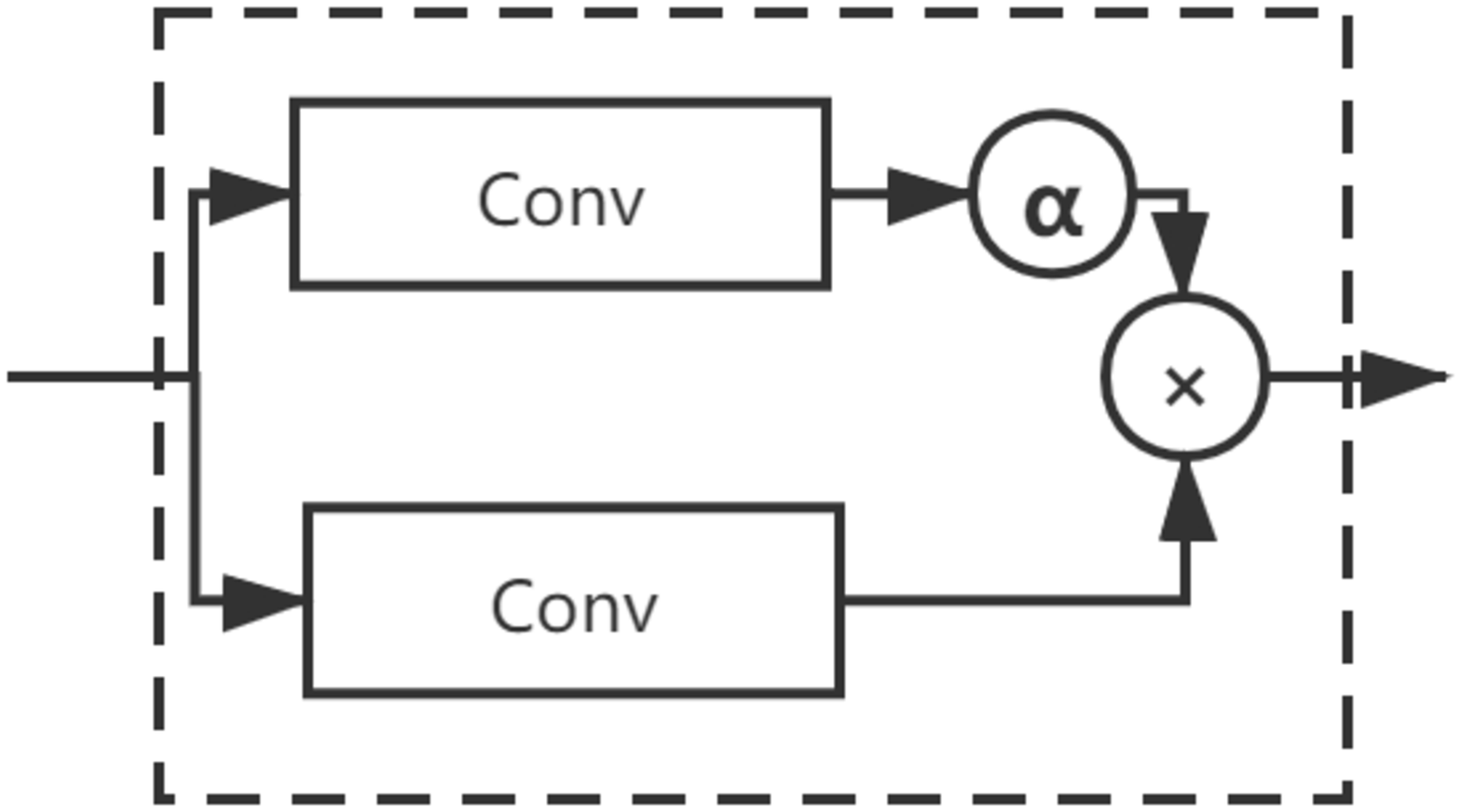}
}
\subfigure[Gated Deconvolution]{
\label{fig:b}
\includegraphics[width=1.3in]{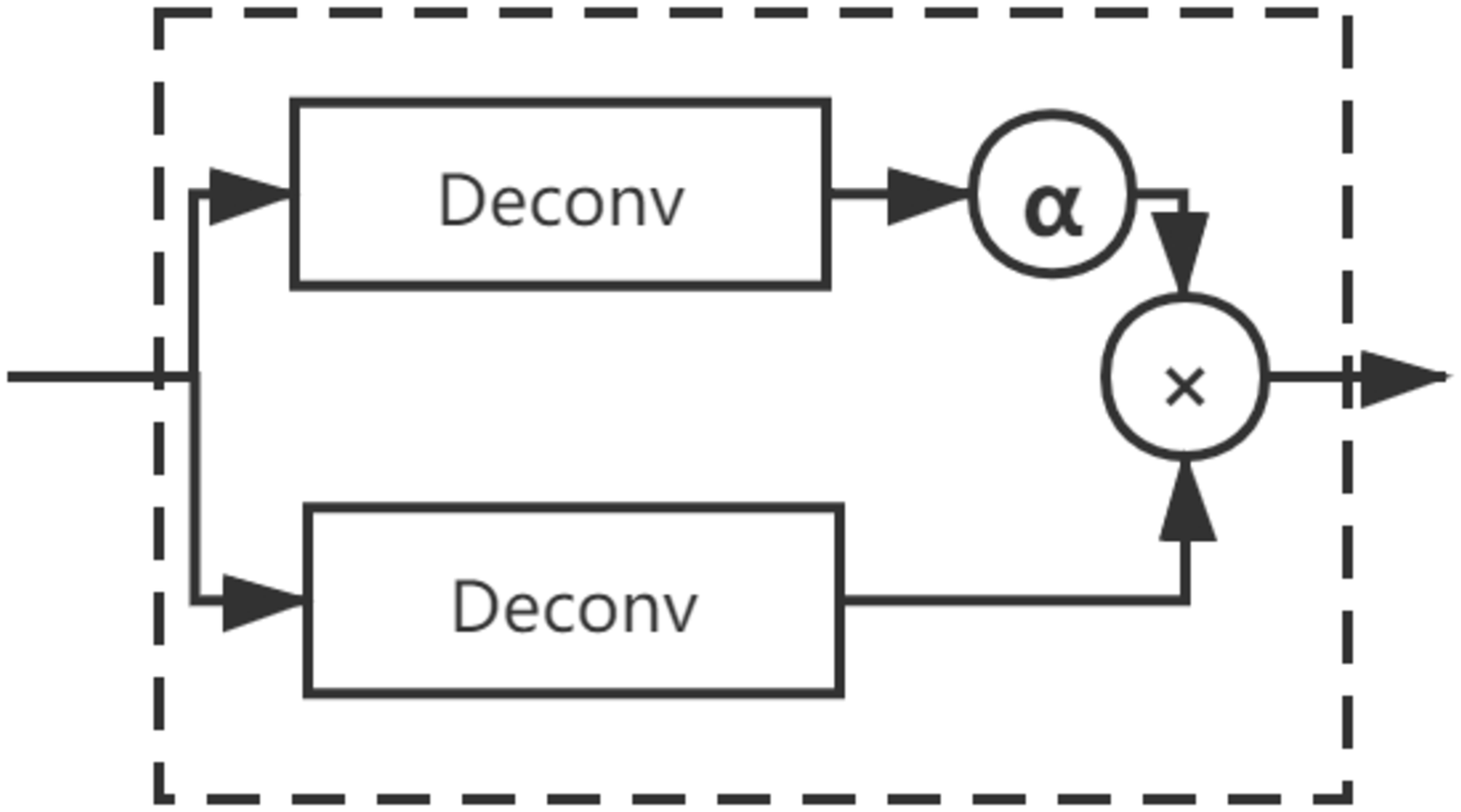}
}
\caption{Gated Convolution and Gated Deconvolution, where $\alpha$ denotes a sigmoid funtion}
\label{fig:GCN}
\end{figure}

    Dauphin et al. ~\cite{GateCNN} improved the masked convolution for image convolution modeling and proposed gated convolution (GCNN) which is described as:
    ~\begin{equation}
    \begin{aligned}
        y &= (x \ast W_1 +b_1) \odot \sigma{(x \ast W_2 +b_2)} \\&=v_1 \odot \sigma{(v_2)} 
    \end{aligned}
    \end{equation}
    where $W$ and $b$ represent kernel and bias, respectively. $\ast$ and $\odot$ denote operation of convolution and element-wise multiplication, respectively. $\sigma$ represents a nonlinear activation function. The GCNN can reduce the vanishing gradient problem for deep architectures by providing a linear path for the gradients, so we replaced the convolution in the original crn with it. A diagram of gated convolution is shown in Fig.~\ref{fig:GCN}.

\begin{figure}[ht]
  \centering
  \includegraphics[width=0.2\textwidth]{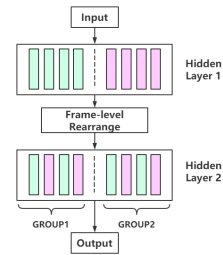}
  \caption{The process of group LSTM}
  \label{fig:GCLSTM}
\end{figure}

    Model efficiency is pretty important, and many application scenarios have higher requirements for processing speed and memory usage. However, due to the introduction of the dual-branch structure, the amount of calculation and memory occupied by the model are much higher than that of the standard convolutional recurrent network. Gao et al.~\cite{gao2018efficient} proposed a grouped recursive neural network (RNN) strategy, which reduces the complexity of the model while ensuring the performance of the model. The process of a group RNN is shown in Fig. \ref{fig:GCLSTM}.

    In this paper, the group LSTM contains two layers of RNN, and each layer has two LSTMs to learn the features within each group. Between the two layers, a frame-level rearrangement is used to establish the inter-group relationship of features, which guarantees the utilization of inter-group correlations to a certain extent.
    

\subsection{Bridge Layer}

    Bridge Layer is a linear unit, which is responsible for converting information from one branch to another. Actually, the bridge layer consists of vectors with the same length as the frame length. We take the real part of the fast Fourier transform (FFT) variable as the initialization parameter of the vectors to fit the situation that SRS is used as the frequency domain representation. 

\subsection{Parameters Setting}
    The encoder has six layers, each consist of a bridge layer, a gated convolution followed by batch normalization and ELU nolinearity. Note that a module is added before the input of frequency branch to calculate the frequency-domain representation of the real number field. The input size of the model is [batch\_size, 1, seq\_len, features]. The first layer of the encoder expands the input channel from 1 to 64. After that,the rest layer of the encoder do the following operations. The bridge layer first translates the features from the other branch and then concatenate it with the output of the previous layer along the channel axis, finally pass the features to a gated convolution. The kernel and the stride of convolution are set to (1, 3) and (1, 2), respectively. The final output size of the encoder is [batch\_size, 64, seq\_len, features / 64]. The output is fed into group LSTM layers to extract the long-term relationship of the features in time domain and frequency domain, respectively.

    Similarly, the decoder also has six layers. In addition to the features from another domain, each layer also gets the skip connection of the corresponding layer of the encoder, and concatenate them with the output of the previous layer along the channel axis. The decoder uses gated deconvolution to double the feature dimensions layer by layer to reconstruct the signal to the original size. The last layer of decoder enhance the signal to one channel. Finally, the output is converted into speech through the overlap-and-add operation. Note that the speech from frequency branch is regarded as final result.

\subsection{Loss Function}
    In the early experiments, we used a loss based on STFT magnitude which was proposed in ~\cite{pandey2019new} and can be described as:
    ~\begin{equation}
    \begin{aligned}
        L_{Mag}(s,\hat{s}) =\frac{1}{T \cdot F}\sum_{t=0}^{T-1}&\sum_{f=0}^{F-1}[(|S_r(t,f)|+|S_i(t,f)|)\\&-(|\hat{S}_r(t,f)|+|\hat{S}_i(t,f)|)]
    \end{aligned}
    \end{equation}
    where $T$ and $F$ represent the number of time frames and frequency dimensions, $S$ and $\hat{S}$ denote STFTs of s and $\hat{s}$, respectively. $S_r$ and $\hat{S}_i$ represent the real and imaginary parts of $S$, respectively. Note that the output of the network contains two enhanced utterances, one of which is from the time branch and the other from the frequency branch and they are optimized independently. So the total loss is defined as:
    ~\begin{equation}
        L=L_{Mag}(s,\hat{s}_t)+L_{Mag}(s,\hat{s}_f)
    \end{equation}
    
    However, we found that magnitude loss introduced a large number of unknown artifacts. Although it does not affect the objective evaluation score, it would bring terrible auditory perception. Therefore, in DNS Challenge, the magnitude loss is replaced with the phase constrained magnitude loss proposed in ~\cite{2020arXiv200901941P} and achieved good subjective evaluation scores in the competition.
\section{Experiment}
\label{Sec:experiment}

\subsection{Datasets}
    In this study, we evaluated the performance of our proposed model on the WSJ0 SI-84 dataset~\cite{WSJ0} which includes 7138 utterances from 83 speakers (42 males and 41 females). We used the utterances of 77 speakers for training and the rest for test. We used 10000 non-speech sounds from a sound effect library (available at \url{www.sound-ideas.com})~\cite{sound-ideas} and generated 320000 and 3000 utterances at the SNRs uniformly sampled from \{-5dB, -4dB, -3dB, -2dB, -1dB, -0dB\} for training and validation, respectively. For the test set, two noises (babble and cafeteria) from Auditec CD (available at \url{http://www.auditec.com}) are used to generate 300 mixtures at each SNR of -5dB, 0dB, and 5dB.
    
\subsection{Baselines}
    In this study, we compared the proposed dual-branch network with another 3 baselines, namely CRN~\cite{CRN}, GCRN~\cite{tan2019learning} and AECNN~\cite{pandey2019new}, which are given as follows:
    
    \begin{itemize}
    \item CRN: it is a casual convolutional recurrent network in T-F domain. The network uses 5 convolution layers as the encoder and 5 deconvolution layers as the decoder. Two LSTM layers are used for sequence modeling. This network receiving magnitude as input. The number of channels is decreased and the number of parameters is 4.5M.
    \item GCRN: it is a causal gated convolutional recurrent network for complex spectral mapping. The structure is similar to CRN except that GCRN has two decoders to model real and imaginary, respectively. The input of network is complex spectral. We kept the best configuration in~\cite{tan2019learning} and the number of parameters is 9.76M.
    \item AECNN: it is a autoencoder based fully convolutional neural network in the time domain. Raw waveform is chunked into frames with a large tiem frame size (1.024s). We kept the best configuration in~\cite{pandey2019new}. The number of parameters is 18M.
    \item DBNet: the structure of two branches are same. 6 (de) convolution block are set for encoder and decoder. The number of channels is 64 for each layer. The kernel size of (1, 3) and stride of (1, 2) are set for time and frequency axis. The input is time frames and SRS for time branch and frequency branch, respectively. The numebr of parameters is 2.9M.
    \end{itemize}

\subsection{System Settings}
    All utterances are sample at 16kHz. The frames are extracted using a rectangular window and a hamming window of size 20ms for time domain and frequency domain, respectively. The overlap is 10ms. The models are trained using the Adam optimizer ~\cite{Adam} with a learning rate of 0.001. And the batch size is set to 32 at the utterance level. Note that a random segment of 7 seconds is intercepted from an utterance if it is larger than 7 seconds. The smaller utterances are zero-padded to match the size of the largest utterance in the batch.
\subsection{Evaluate Metrics}
    The performance is evaluated with two objective metrics: short-time objective intelligibility (STOI)~\cite{STOI} and perceptual evaluation of speech quality (PESQ)~\cite{PESQ}. STOI values typically range from 0 to 1, which can be roughly interpreted as percent correct. PESQ values range from -0.5 to 4.5. For both of the STOI and PESQ metrics, the higher number indicates better performance.

\section{Result and Analysis}
\label{Sec:result}
 
 \begin{table*}[ht]
\centering
\caption{STOI and PESQ comparisons between DBNet and the baseline models}
\label{tab:wsj0_result}
\resizebox{\textwidth}{!}{
\begin{tabular}{|c|l|l|l|l|l|l|l|l|l|l|l|l|l|l|l|l|c|}
\hline
\multicolumn{1}{|c|}{Metrix} & \multicolumn{8}{c|}{STOI} & \multicolumn{8}{c|}{PESQ} &
\multirow{4}{*}{Casual?} \\ \cline{1-17}
\multicolumn{1}{|c|}{Test Noise} & \multicolumn{4}{c|}{Babble} & \multicolumn{4}{c|}{Cafeteria} & \multicolumn{4}{c|}{Babble} & \multicolumn{4}{c|}{Cafeteria} & \\ \cline{1-17}
\multicolumn{1}{|c|}{Test SNR} & \multicolumn{1}{c|}{-5db} & \multicolumn{1}{c|}{0db} & \multicolumn{1}{c|}{5db} & \multicolumn{1}{c|}{AVG} & \multicolumn{1}{c|}{-5db} &
\multicolumn{1}{c|}{0db} & \multicolumn{1}{c|}{5db} & \multicolumn{1}{c|}{AVG} &
\multicolumn{1}{c|}{-5db} & \multicolumn{1}{c|}{0db} & \multicolumn{1}{c|}{5db} &
\multicolumn{1}{c|}{AVG} & \multicolumn{1}{c|}{-5db} & \multicolumn{1}{c|}{0db} &
\multicolumn{1}{c|}{5db} & \multicolumn{1}{c|}{AVG} & \\ \cline{1-17}
\multicolumn{1}{|c|}
{Mixture} &
  59.1 & 71.6 & 83.3 & 71.3 & 57.5 & 70.1 & 82.4 & 70.0 &
  1.52 & 1.76 & 2.07 & 1.78 & 1.40 & 1.69 & 2.07 & 1.72 &
   \\ \hline
CRN &
  77.9 & 88.0 & 93.2 & 86.4 & 75.7 & 86.6 & 92.7 & 85.0 &
  1.99 & 2.50 & 2.91 & 2.47 & 2.01 & 2.47 & 2.89 & 2.46 &
  \checkmark \\ \hline
GCRN &
  81.1 & 90.7 & 94.7 & 88.8 & 78.0 & 88.8 & 94.1 & 87.0 &
  2.03 & 2.65 & 3.07 & 2.58 & 1.97 & 2.56 & 3.03 & 2.52 &
  \checkmark \\ \hline
AECNN &
  80.5 & 90.6 & 94.2 & 88.4 & \textbf{80.0} & 89.4 & 94.0 & 87.8 &
  2.00 & 2.57 & 2.88 & 2.48 & 2.03 & 2.55 & 2.93 & 2.50 &
  \checkmark \\ \hline 
DBNet &
  \textbf{83.1} & \textbf{91.8} & \textbf{95.3} & \textbf{90.0} & 
          79.6  & \textbf{89.9} & \textbf{94.8} & \textbf{88.1} &
  \textbf{2.15} & \textbf{2.78} & \textbf{3.16} & \textbf{2.70} & 
  \textbf{2.08} & \textbf{2.69} & \textbf{3.10} & \textbf{2.62} &
  \checkmark \\ \hline \hline
GCRN-NC &
  84.1 & 92.1 & 95.5 & 90.5 & 81.3 & 90.5 & 95.0 & 89.0 &
  2.22 & 2.81 & 3.17 & 2.73 & 2.08 & 2.72 & 3.07 & 2.62 &
  × \\ \hline 
DBNet-NC &
  \textbf{87.3} & \textbf{93.5} & \textbf{96.2} & \textbf{92.3} & 
  \textbf{84.3} & \textbf{91.9} & \textbf{95.7} & \textbf{90.6} &
  \textbf{2.56} & \textbf{3.06} & \textbf{3.34} & \textbf{2.99} & 
  \textbf{2.43} & \textbf{2.95} & \textbf{3.31} & \textbf{2.90} &
  × \\ \hline
\end{tabular}}
\end{table*}      
\subsection{Objective Comparisons}
    First, we compare our model with baselines in terms of both STOI and PESQ. The results are given in Table \ref{tab:wsj0_result} and the best results are marked in bold. For STOI, DBNet is better than all baselines at all SNRs and noises except AECNN. However, AECNN is a model based on large frame size so it is not suitble for real-time scenes. Compared with GCRN, an average improvement of 1.20\% and 1.10\% is obtained for babble and cafeteria, respectively. For PESQ, GCRN is the best baseline and an average improvement of 0.12 and 0.10 is obtained for babble and cefeteria, respectively. As for non causal system, DBNet-NC outperforms GCRN-NC and obtained an average improvement of 1.7 for STOI and 0.27 for PESQ.

    In conclusion, the proposed dual-branch model outperforms both AECNN which is a time-domain based model and GCRN which is a frequency-domain based model for complex spectrogram mapping, indicating that alternate interconnection of the information of two domains can significantly improve the performance of the model and improve parameter utilization.

\subsection{Deep Noise Suppression Challenge}
    The Deep Noise Suppression (DNS) challenge is designed to foster innovation in the area of noise suppression to achieve superior perceptual speech quality. To evaluate the performance of the proposed method on more complicated and real acoustic scenarios, the proposed model was trained with the DNS-Challenge wide band dataset which contains more complex acoustic scenarios including reverberation, singing, emotions, and non-English speech. The settings of generating training set is described as follow. The SNRs of training mixtures vary from -5dB to 25dB. Around 30\% of utterances are convolved with provided synthetic and real room impulse responses (RIRs) before mixed with different noise signals, and we process speech, noise and reverberation by using the spectral augmentation filters proposed in ~\cite{2017A}. Moreover, there is a 5\% chance that there may be multiple compound noises in one utterance.
    To meet the requirement of DNS-Challenge, the number of channels is appropriately decreased. In addition, kernel size of convolution blocks is set to (2, 3).

    We use DNSMOS~\cite{reddy2020dnsmos} which is a reliable non-intrusive objective speech quality metric as our evaluation metrics at training stag and take Mean Opinion Score (MOS) of the ITU-T P.835 framework as result. The results of the evaluation using the ITU-T P.835 criterion \cite{naderi2020crowdsourcing} which is provided by the organizer are shown in Table~\ref{tab:dns_result}. It is obvious that the proposed model outperforms the baseline (NSnet2) by overall 0.12 DMOS score. Then we calculated the processing latency of our algorithm according to the competition requirements. In this model, the frame size T = 20ms, and the overlap between consecutive frames Ts = 10ms, so the latency T =30ms, which meets the requirements. We also evaluated the memory access cost (MAC), and the result is 2.847G per second.

\begin{table}[ht]
\caption{Subjective evaluation with P.835 criterion on the DNS Challenge}
\label{tab:dns_result}
\resizebox{0.47\textwidth}{0.6cm}{
\small
\begin{tabular}{cccccccc}
\hline
       & Stationary & Emotional & Tonal & Non-English & Musical & English & Overall \\ \hline
Noisy  & 3.03       & 2.28      & 3     & 3.04        & 2.57    & 2.52    & 2.77    \\
NSnet2 & 3.28       & 2.75      & 3.31  & 3.25        & 2.78    & 2.93    & 3.07    \\ 
DBNet  & 3.44       & 3.07      & 3.37  & 3.43        & 2.78    & 2.93    & 3.19    \\ \hline
\end{tabular}}
\end{table}

\section{Conclusions}
\label{Sec:conclusions}
    In this study, we propose a novel single-channel speech enhancement system, which consists of two denoising branches on time domain and frequency domain. The results turns out that the proposed model outperforms other advanced models in terms of objective intelligibility and quality scores. 

    We explain our work has excellent performance because the information in the time domain and the frequency domain is not exactly the same. According to the principle of STFT, convolution in the time domain is equivalent to direct product in the frequency domain. Operations in the time domain tend to focus more on local information, and operations in the frequency domain focus more on the relationship between frames. A reasonable combination of the two can achieve better performance. And the proposed model has fewer parameters which indicate the two branch structure improves parameter utilization. Subjective results showed that the proposed system ranks the top 8 in the Mean Opinion Score (MOS) of the ITU-T P.835 for real-time track 1 of the INTERSPEECH 2021 Deep Noise Suppression (DNS) challenge. 

\section{Acknowledgements}
The author would like to thank Yongjie Yan, Tailong Zhang and Pengjie Shen for their valuable comments. This research was partly supported by the China National Nature Science Foundation (No. 61876214).
\bibliographystyle{IEEEtran}
\clearpage
\bibliography{mybib}

\end{document}